\documentclass{article}

     \usepackage[preprint,nonatbib]{neurips_2019}

\usepackage[utf8]{inputenc} % allow utf-8 input
\usepackage[T1]{fontenc}    % use 8-bit T1 fonts
\usepackage{hyperref}       % hyperlinks
\usepackage{url}            % simple URL typesetting
\usepackage{booktabs}       % professional-quality tables
\usepackage{amsfonts}       % blackboard math symbols
\usepackage{nicefrac}       % compact symbols for 1/2, etc.
\usepackage{microtype}      % microtypography
\usepackage[compact]{titlesec} 
\titlespacing{\subsection}{5pt}{*1}{*1}
\usepackage[skip=2pt]{caption}
\usepackage[pdftex]{graphicx}

\title{Combining human cell line transcriptome analysis and Bayesian inference to build trustworthy machine learning models for prediction of animal toxicity in drug development}

\author{%
  Laura-Jayne Gardiner \\
  IBM Research\\
  The Hartree Centre, UK\\
  Laura-Jayne.Gardiner@ibm.com\\
  \And
  Anna Paola Carrieri\\
  IBM Research\\
  The Hartree Centre, UK\\
  acarrieri@uk.ibm.com\\
  \And
  Jenny Wilshaw\\
  IBM Research\\
  The Hartree Centre, UK\\
  jwilshaw@rvc.ac.uk\\
  \And
  Stephen Checkley\\
  STFC Daresbury Lab\\
  Warrington, UK\\
  Stephen.checkley@stfc.ac.uk\\
  \And
  Edward O. Pyzer-Knapp\\
  IBM Research\\ 
  The Hartree Centre, UK\\
  epyzerk3@uk.ibm.com\\
  \And
  Ritesh Krishna\\
  IBM Research\\
  The Hartree Centre, UK\\
  Ritesh.krishna@uk.ibm.com\thanks{Corresponding Author}
}

\begin{document}

\maketitle

\begin{abstract}
Biomedical data, particularly in the field of genomics, has characteristics which make it challenging for machine learning applications – it can be sparse, high dimensional and noisy. Biomedical applications also present challenges to model selection – whilst powerful, accurate predictions are necessary, they alone are not sufficient for a model to be deemed useful. Due to the nature of the predictions, a model must also be trustworthy and transparent, empowering a practitioner with confidence that its use is appropriate and reliable. In this paper, we propose that this can be achieved through the use of judiciously built feature sets coupled with Bayesian models, specifically Gaussian processes. We apply Gaussian processes to drug discovery, using inexpensive transcriptomic profiles from human cell lines to predict animal kidney and liver toxicity after treatment with specific chemical compounds. This approach has the potential to reduce invasive and expensive animal testing during clinical trials if \textit{in vitro} human cell line analysis can accurately predict model animal phenotypes. We compare results across a range of feature sets and models, to highlight model importance for medical applications.  
\end{abstract}

\section{Drug Discovery Challenge Description}

To progress a pharmaceutical drug candidate to human trial phase, regulations require preclinical trials or animal tests to ascertain compound safety. This is balanced against a desire to reduce animal testing of compounds due to expense and ethical implications. However, although compound toxicity in animals can translate to human toxicity, this translation shows considerable variability that we need to understand \cite{Bailey1, Bailey2, Poussin}. Machine learning (ML) offers a potential path to derive insight on this complex task. To enable this, there has been a large amount of effort invested in cost effective techniques to generate data-sets from which the likely effect that a compound will have after administration, particularly with respect to toxicity or adverse events, can be predicted. One such approach is L1000, a high-dimensional gene expression profiling method, which is fast emerging as a cheap alternative to produce large amounts of experimental observations suitable for ML driven drug discovery \cite{Subramanian}. 

We use L1000 profiles generated for a variety of human cell lines under different perturbation conditions for a large number of chemical compounds. We investigate the use of L1000 profiles to generate predictive models and the validity of their application on rats – a model drug testing species. We apply ML to L1000 profiles generated after chemical compound treatment, to predict rat kidney dysfunction after treatment with the same compound. By including feature information about the chemical compounds used to treat human cells and animals during preclinical trials, we translate these inexpensive and non-invasive human cell line tests to preclinical phenotypic responses in model species such as rats. L1000 gene expression profiles and chemical structures, were collated by \cite{Wang} including 964 “Landmark” genes following small molecule treatment. L1000 gene expression information is available as -1, 0 and 1 values to represent down-regulation, no change and up-regulation of gene expression. Chemical structure of the compounds was in the form of a 166-bit MACCS chemical fingerprint for small molecule compounds computed using Open Babel \cite{OBoyle}. 

The target for our ML approach is the prediction, in rats, of the amount of blood urea nitrogen (BUN). Elevated BUN tests are routinely used as an indicator that the kidneys or liver may be damaged or dysfunctional. In rats elevated levels for BUN of 28-136mg/dl were consistent with kidney dysfunction \cite{Stender, Yan}. The datasets used here were collected from DrugMatrix and detail the rat blood concentration of BUN after compound administration. BUN levels range from 7.75-39.60mg/dl encompassing a wide range from known safe levels to high levels that exceed health indicator limits. We matched the induced L1000 gene expression signatures with rat blood test results where the same compound had been used. This allowed us to match L1000 profiles with 429 rat BUN measurements. These observations all represent different chemical compound treatments for the training set.

We predict BUN levels using either human gene expression data or compound chemical structure information as features. Then, we combine both feature sets (1,130 features) and show substantial prediction improvement in line with previous work \cite{Wang}. We use dimensionality reduction to tackle problems associated with sparse and high dimensionality datasets. We highlight the benefit of using Gaussian processes to understand the transparency and trustworthiness of a prediction, which is of high importance in healthcare. We integrate heterogeneous public experimental datasets and, for the first time, we translate predictive capability related to renal toxicity or BUN level from \textit{in vitro} human cell line tests to \textit{in vivo} blood tests from model animal species. Our ultimate aim is to reduce expensive and invasive animal testing with the ability to predict clinical response using human cell line gene expression profiles and mathematical modelling, supporting the pharmaceutical industry’s commitment to the 3R’s (Replacement, Reduction, Refinement) in drug development \cite{Parker}.

\section{Approach and Results}

\subsection{Dataset enrichment through augmentation of L1000 features with chemical structure}
A richer feature set, coupled with an appropriate model, should improve the predictive power of the approach. As such, we augmented L1000 features with chemical structure information. To test this, we applied Gaussian Process (GP) regression \cite{Pedregosa, Rasmussen} to predict BUN level using firstly, only the L1000 gene expression profiles as the training dataset (964 features). Secondly, using only chemical structure information as the training dataset (166 features). Finally, we combined both information resources gene expression and chemical structure to generate a single training dataset (1,130 features). For all three analyses we used a kernel which was the sum of a simple RBF kernel, and a WhiteKernel. Hyperparameters were optimized using descent on the marginal log-likelihood of the model.  Due to the large dimensionality of the data, a single length scale was used over all features. 

\begin{figure}[ht!]
  \centering
  \includegraphics[width=0.9\linewidth]{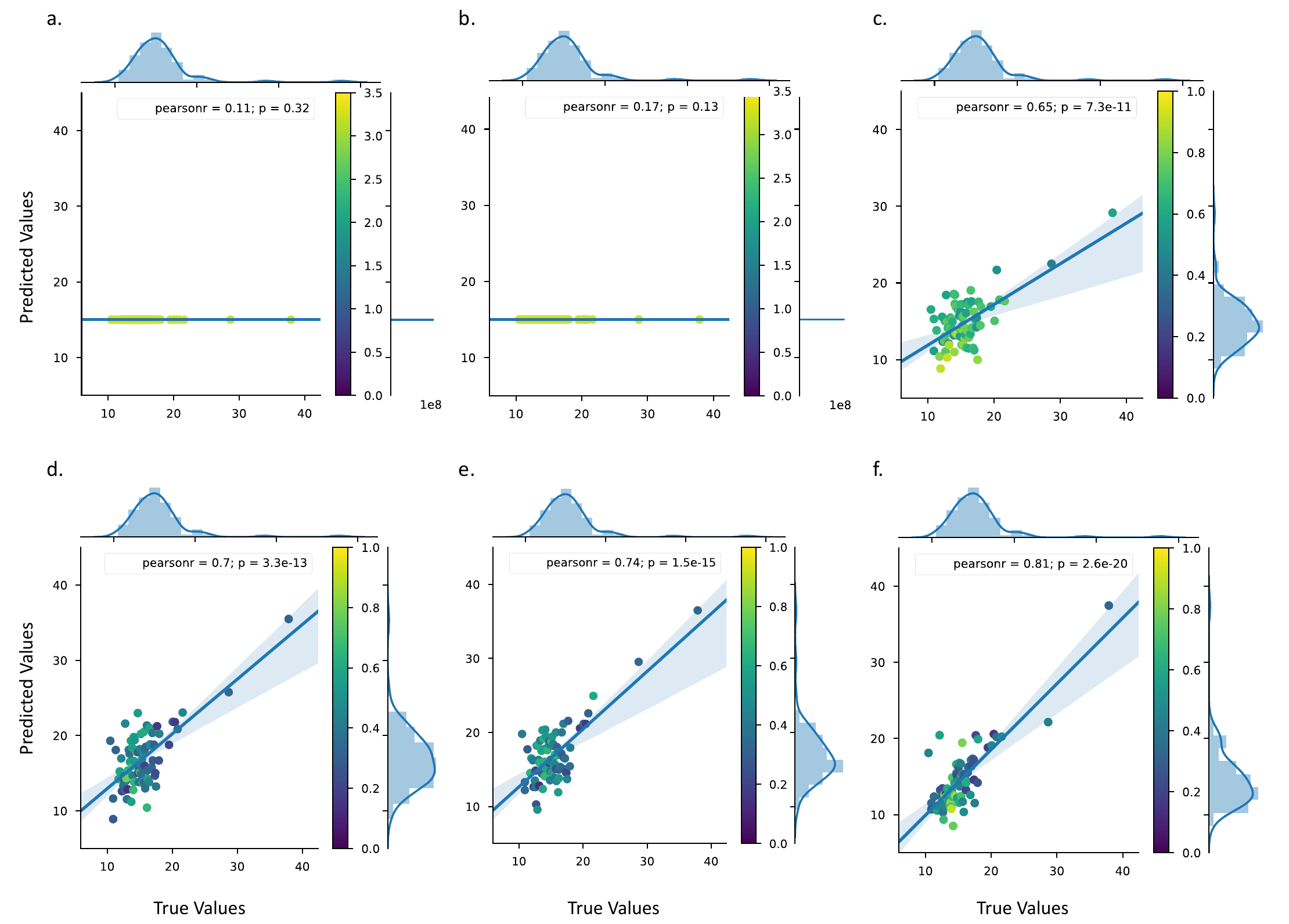}
  \caption{ True (x-axis) vs predicted (y-axis) test values for BUN level prediction: using as training data (a) L1000 gene expression, (b) chemical structure information and (c) both; using gene expression and chemical structure training data after (d) PCA, (e) tSVD and (f) tSVD h. Plots show marginal histograms with regression and kernel density fits. The 95\% confidence interval for regression is drawn using translucent bands around the regression line. Point colour is based on the standard deviation of its underlying predictive distribution: note that point colour scales vary between plots.
\label{fig:bun_data_comparison}}
\end{figure}

Figure~\ref{fig:bun_data_comparison} and Table 1 show that the combination of gene expression and chemical structure information decreases test set RMSE scores to 2.961 from the average 3.708 seen for chemical structure and gene expression separately. There is also a marked increase in test set r2 scores after combination of the features from on average 0.021 to 0.418. Figure~\ref{fig:bun_data_comparison} shows the true versus predicted values for the test dataset where, correlation coefficients increase from 0.11 and 0.17 for gene expression and chemical structure, to 0.65 for combined features. In addition, Figure~\ref{fig:bun_data_comparison} shows that confidence in the predictions increases after feature combination with the standard deviation of the predictive distribution for each of the test datapoints showing an average decreasing from 3.283 and 3.283 for gene expression and chemical structure respectively to 0.683 for the combined features. The combination of gene expression and chemical structure information increases our predictive capability for BUN level. This analysis also highlights the benefit of using a Gaussian Process with the ability to directly capture the uncertainty of the predictions. We observe a dramatic decrease (4.8-fold) in the uncertainty level or the standard deviation of the predictive distribution of the predictions after feature combination while the difference in RSME scores was not so marked (1.3-fold) and could mask a poor model (Table 1). 

\begin{table}[ht!]
\centering
\tiny
\begin{tabular}{|c|c|c|c|c|c|c|c|}
\hline
& \textit{Gene Expr.} 
& \textit{Chem.}   
& \multicolumn{5}{c|}{\textit{Gene Expression \& Chemical Structure}} 
\\ \hline

\multicolumn{1}{|l|}{} & \textbf{GP}                                               & \textbf{GP}    
                        & \textbf{GP}    
                         & \textbf{GP\&PCA}             
                         & \textbf{GP\&TSVD}
                         & \textbf{GP\&TSVD h.}
                         & \textbf{LightGBM}            
                         \\ \hline

\textbf{Train MAE}     & 1.988                                                   & 1.988                                                   & 0                                                       & 0                                                       & 0                                                       & \textbf{0}                                                       & 
\begin{tabular}[c]{@{}c@{}}0.501\\ %**3.334 $\pm{1.060}$
\end{tabular} \\ \hline
\textbf{Test MAE}      & 2.156                                                   & 2.156                                                   & 2.365                                                   & 2.7                                                     & 2.452                                                   & \textbf{1.71}                                                    & 1.544                                                          \\ \hline
\textbf{Test RMSE}     & \begin{tabular}[c]{@{}c@{}}3.708\\ *3.708\end{tabular} & \begin{tabular}[c]{@{}c@{}}3.708\\ *3.708\end{tabular} & \begin{tabular}[c]{@{}c@{}}2.961\\ *2.240\end{tabular} & \begin{tabular}[c]{@{}c@{}}3.359\\ *2.157\end{tabular} & \begin{tabular}[c]{@{}c@{}}3.176\\ *2.047\end{tabular} & \begin{tabular}[c]{@{}c@{}} \textbf{2.419}\\ \textbf{*1.528}\end{tabular} & 2.146                                                          \\ \hline
\textbf{Test r2}       & 0.013                                                   & 0.028                                                   & 0.418                                                   & 0.464                                                   & 0.52                                                    & \textbf{0.661}                                                   & 0.656                                                          \\ \hline
\textbf{Av. SD}        & 3.283                                                   & 3.283                                                   & 0.683                                                   & 0.427                                                   & 0.425                                                   & \textbf{0.47}                                                    & -                                                              \\ \hline
\end{tabular}
\smallskip
\caption{Prediction of BUN levels. Best results from comparison of combining chemical and gene expression data, using dimensionality reduction and finally comparing the best alternative classifier from 8 tested, defined as producing the highest test prediction accuracy (lowest MAE with highest r2 score). The best model is highlighted in bold. *Weighted RMSE.
\label{table}}
\end{table}

\subsection{Dealing with high dimensional data}

One of the main challenges for this data was a greater number of features than datapoints (1,130 vs 429) as traditional ML methods could overfit. To address this, firstly, we use a non-parametric Bayesian model known as a Gaussian process (GP) that uses a covariance matrix, rather than the input features themselves, so its predictive power scales with datapoint number, rather than dimensionality. We found a kernel comprising of a sum of the common Radial Basis Function (RBF) kernel, with automatic relevance determination (ARD) to best model the data. Secondly, we used dimensionality reduction methods that are useful in building ML models from genomics datasets \cite{Trapnell, Luecken, rowe} implementing a truncated singular value decomposition (t-SVD) approach \cite{Hansen, Eckart}.  We implemented a hierarchical approach to t-SVD (tSVD h.), where the two parts of the feature space (L1000 profiles and chemical descriptors) were considered separately and then recombined.  Preserving 95\% of feature information per sub-domain resulted in a 57-dimensional vector, with 17 dimensions representing chemical features and 40 dimensions representing L1000 profiles. We tested tSVD h. against a straight dimensionality reduction to 57 dimensions using t-SVD and the commonly used PCA method (Figure~\ref{fig:bun_data_comparison}, Table 1). Our tSVD h. approach gave the strongest results with a strong correlation coefficient of 0.81 and r2 of 0.661 exceeding any other from this dataset and the lowest weighted RMSE (1.528). All methods outperform a GP with no dimensionality reduction.

\subsection{Capturing Model Uncertainty to Improve Trust and Transparency}

In healthcare, one of the barriers to the uptake of ML is a perceived lack of trust. We believe that a lack of trust originates mainly from poor prediction on out of set problems and a lack of transparency/interpretability. To address poor out of set modelling, we propose building a rich description of model uncertainty and returning both a prediction and confidence on new data. Thus, an out of set prediction will come with a ‘warning’ about possible performance issues. To demonstrate the importance of predictive uncertainties, we fitted commonly used deterministic methods to the data – Linear regression, Random Forest, Support Vector Regressor, XGBoost, Gradient Boosting, K nearest neighbors and LightGBM. 80\% of the data was used for training and the remaining 20\%, with no overlap with the training set, was held out for testing. 10-fold cross validation was performed on the training data (ShuffleSplit) and hyperparameters tuned using a grid search. We compared results to our best GP model using the best model for each method. GP performed favourably to other algorithms with its closest rival model generated with LightGBM (Table 1). However, the inclusion of uncertainty measurements with GP allows calculation of a weighted RSME. The low uncertainty rate (0.470) from the GP results in the lowest observed RMSE (1.528) across all of the models after weighting and the highest r2 score of 0.661 (Table 1). Finally, we highlight a specific sample in the top right corner of Figure~\ref{fig:bun_data_comparison}~f with a BUN level of 37.89mg/dl consistent with potentially severe kidney dysfunction. The LightGBM model predicts the BUN level of this sample as 34.61mg/dl decreasing it by 3.3mg/dl. GP predicts 37.23mg/dl representing a 5-fold prediction improvement. Furthermore, the GP's low uncertainty estimate of 0.661 confirms the close proximity of the prediction to the true value. Here, the absence of this information for LightGBM, could result in severe kidney dysfunction that may need immediate intervention, being mis-diagnosed for a more moderate condition. 

We can improve trust in a model by generating insight into the contributing factors to a prediction. We exploit the fact that the length scales of our GP are fitted per feature and can give insight into the impact specific features have on our model – shorter length scale, higher impact. However, since we built our features using t-SVD, we use an inverse projection to relate the reduced feature space to the original features for the dimension, focusing on the shortest length scale in our best GP model based on gene expression. The top three genes that represent this reduced feature space have relevant roles and include: CCL2 that is linked to renal damage \cite{Kashyap}, FOSL1 that regulates cell survival and GLRX that is expressed in the kidney and contributes to the defense system. This analysis highlights important genes that are closely associated with the biological purpose of the model. Using length scale in this way, therefore improves the transparency and trustworthiness of the GP model. This analysis outperformed the ExtraTreesRegressor approach that we trialed for the same purpose. This method does not consider the model itself and therefore it defined genes that were less biologically relevant e.g. TUBB6 (tubulin beta 6 class V) and DAXX (Death Domain Associated Protein).

\section{Conclusions}

For ML in healthcare, there are challenges emanating from the data (sparsity, low volume, lack of structure) or the user (requirements for transparency and trust). We use GP to confidently predict BUN level in rats after compound treatment using gene expression profiling of human cell lines. We demonstrate transferability of information between human and animal subjects in the effort to reduce the need for animal testing and eliminate toxic compounds earlier in the evaluation process to reduce development costs. We use t-SVD to reduce the high dimensional input data, which displays superior predictive power when used with a GP regressor. This reduced dimensionality allows us to improve the quality of the predictive uncertainties, through the addition of feature-wise length scales to the kernel. We also fit a noise kernel to the data to account for some non-correlated noise in the measurements of toxicity. The predictive uncertainties recovered from the GP improve the transparency and trustworthiness of the model, allowing the end users to understand when it can and cannot be used.  Finally, we further improve the transparency providing insight into the feature-wise contribution to the prediction – a starting point for explainable AI for medical applications.

\subsubsection*{Acknowledgments}

This work was supported by the STFC Hartree Centre’s Innovation Return on Research programme, funded by the Department for Business, Energy \& Industrial Strategy.

%\section*{References}

%References


\begin{thebibliography}{9}
\bibitem{Bailey1}
Bailey, J., Thew, M. \ \& Balls, M.\ (2014) An Analysis of the Use of Animal Models
 in Predicting Human Toxicology and Drug Safety.
 {\it Alternatives to Laboratory Animals} {\bf 42}(3):181–199.

\bibitem{Bailey2} 
 Bailey, J., Thew, M. \ \& Balls, M. \ (2013) An Analysis of the Use of Dogs in 
 Predicting Human Toxicology and Drug Safety. 
 {\it Alternatives to Laboratory Animals} {\bf 41}(5):335–350.
 
 \bibitem{Poussin}
Poussin, C., Mathis, C., Alexopoulos, L., Messinis, D., Dulize, R. {\it et al.} \ (2014) The 
species translation challenge - A systems biology perspective on human and rat bronchial 
epithelial cells. {\it Scientific Data } {\bf 1}, Article number: 140009.

\bibitem{Subramanian}
Subramanian, A., Narayan, R., Corsello, S.M., Peck, D.D., Natoli, T.E. {\it et al.} \ (2017) 
A Next Generation Connectivity Map: L1000 Platform and the First 1,000,000 Profiles. 
{\it Cell} {\bf 171}(6):1437-1452.

\bibitem{Wang}
Wang, Z., Clark, N.R. \ \& Ma’ayan A. \ (2016) Drug-induced adverse events prediction with 
the LINCS L1000 data. {\it Bioinformatics} {\bf 32}(15):2338-2345.

\bibitem{OBoyle}
O'Boyle, N.M. , Banck, M., James, C.A., Morley, C., Vandermeersch, T. \ \& Hutchison, G.R. \ (2011) 
Open Babel: An open chemical toolbox. {\it J. Cheminformatics} {\bf 7}(3):33.

\bibitem{Stender}
Stender, R.N., Engler, W.J., Braun. T.M. \ \& Hankenson, F.C. \ (2007) Establishment 
of blood analyte intervals for laboratory mice and rats by use of a portable clinical analyzer. 
{\it J Am Assoc Lab Anim Sci} {\bf 46}(3):47-52.

\bibitem{Yan}
Yan, S.L., Lin, P. \ \& Hsiao, M. \ (1999) Separation of Urea, Uric Acid, Creatine, and 
Creatinine by Micellar Electrokinetic Capillary Chromatography with Sodium Cholate. 
{\it Journal of Chromatographic Science} {\bf 37}(2):45-50.

\bibitem{Parker}
Parker, R.M. \ \& Browne, W.J. \ (2014) The 
Place of Experimental Design and Statistics in the 3Rs. {\it ILAR Journal} {\bf 55}(3):477-485.

\bibitem{Pedregosa}
Pedregosa, F., Variqueux, C., Gramfort, A., Michel, V., Thirion, B. {\it et al.} \ (2011) Scikit-learn: 
Machine Learning in Python.  {\it Journal of Machine Learning Research} {\bf 12}:2825-2830.  

\bibitem{Rasmussen}
Rasmussen, C.E. \ \& Williams, C. \ (2006) Gaussian Processes for Machine Learning.  {\it MIT Press}. 

\bibitem{Trapnell}
Trapnell, C. \ (2015) Defining cell types and states with single-cell genomics. {\it Genome Research} {\bf 25}:1491-1498.

\bibitem{Luecken}
Luecken, M.D. \ \& Theis, F.J. \ (2019) Current best practices in single-cell RNA-seq analysis: a tutorial. {\it Mol. System Biology} {\bf 15}(6):e8746.

\bibitem{rowe}
Rowe, W.P.M., Carrieri, A.P., Alcon-Giner, C., Caim, S., Shaw, K. {\it et al.} \ (2019) 
Streaming histogram sketching for rapid microbiome analytics. {\it Microbiome} {\bf 7}(1):40.


\bibitem{Hansen}
Hansen, P. C. \ (1987) The truncated SVD as a method for regularization. {\it BIT} {\bf 27}(4): 534–553.

\bibitem{Eckart}
Eckart, C. \ \& Young, G. \ (1936) The approximation of one matrix by another of lower rank. {\it Psychometrika} {\bf 1}(3):211–218. 

\bibitem{Kashyap}
Kashyap, S., Osman, M., Ferguson, C., Nath, M., Roy, B. {\it et al.} \ (2018) Ccl2 deficiency 
protects against chronic renal injury in murine renovascular hypertension. {\it Scientific reports} {\bf 8}:8598.

\end{thebibliography}
\end{document}